\documentstyle[prd,preprint,aps,amsmath,amssymb,graphicx,psfrag,wrapfig,subeqnarray]{revtex}
\begin{document}

\tightenlines

\title{Almost-zero-energy Eigenvalues of\\Some Broken Supersymmetric Systems}

\author{Min-Young Choi\footnote{Email address: witt@phya.snu.ac.kr}
, Choonkyu Lee\footnote{Email address: cklee@phya.snu.ac.kr}}

\address{School of Physics and Center for Theoretical Physics
\\Seoul National University, Seoul 151-742, Korea}

\maketitle

\begin{abstract}
For a quantum mechanical system with broken supersymmetry, we
present a simple method of determining the ground state when the
corresponding energy eigenvalue is sufficiently small. A concise
formula is derived for the approximate ground state energy in an
associated, well-separated, asymmetric double-well-type potential.
Our discussion is also relevant for the analysis of the fermion
bound state in the kink-antikink scalar background.
\end{abstract}

\vfill


\newpage





\newcommand{\be}{\begin{equation}}
\newcommand{\ee}{\end{equation}}
\newcommand{\bea}{\begin{eqnarray}}
\newcommand{\eea}{\end{eqnarray}}
\newcommand{\ba}{\begin{array}}
\newcommand{\ea}{\end{array}}
\newcommand{\bra}{\langle}
\newcommand{\ket}{\rangle}

%

\setcounter{equation}{0}

Supersymmetry(SUSY) and its breaking are fundamental issues in
theoretical particle physics. There have also been numerous
applications of SUSY to quantum-mechanical potential problems
\cite{Witt,Junk}, based on the observation that the spectrum of
the Hamiltonian
\be H_+ = -\frac{d^2}{dx^2} + V_+ (x)~,~~~~~V_+ (x)=W^2 (x) +
W'(x)~~\ee
($W(x)$ is the superpotential, and we set $\hbar = 2m=1$) is
related through SUSY to that of the partner Hamiltonian
\be H_- = -\frac{d^2}{dx^2} + V_- (x)~,~~~~~V_- (x)=W^2 (x) -
W'(x)~. \ee
This formalism has provided us with a number of exactly solvable
quantum mechanical systems for which energy eigenvalues and
eigenfunctions can be found in closed forms. The key properties
that made such feat possible are unbroken SUSY, manifested by the
vanishing energy for the ground state of $H_-$ (or $H_+$), and
shape invariance under the change of parameters for the given
potentials \cite{Gend}-\cite{CKS}. This approach can sometimes be
extended to parameter ranges corresponding to (spontaneously)
broken SUSY \cite{Dutt2,Gang}. But, with SUSY broken, the ground
state energy is no longer equal to zero and this jeopardizes the
possibility of obtaining exact analytic results by the SUSY-based
method in a crucial way.

In this work we will show that, in some broken SUSY case for which
the lowest energy $\bar E(>0)$ for the Hamiltonian $H_+$ or $H_-$
is sufficiently small, a simple perturbative scheme leading to an
easy evaluation of $\bar E$ can be developed. Our discussion finds
useful application in studying the almost-zero-energy fermion
modes in the background of a soliton-antisoliton pair.

The superpotential relevant for our discussion is given as
follows. Let $\sigma_R (x)$ be a generic function with the
properties
\be \begin{split}  \sigma_R (x)>0~~~~~&,~~{\rm for}~x>0~{\rm
and}~|x|~{\rm not}~{\rm very}~{\rm small}~, \\  \sigma_R (x)
\rightarrow -v~~~~~&,~~{\rm for}~x<0~{\rm and}~|x|~{\rm not}~{\rm
very}~{\rm small} \end{split} \ee
and $\sigma_L (x)$ the one with the properties
\be \begin{split}  \sigma_L (x)>0~~~~~&,~~{\rm for}~x<0~{\rm
and}~|x|~{\rm not}~{\rm very}~{\rm small}~, \\  \sigma_L (x)
\rightarrow -v~~~~~&,~~{\rm for}~x>0~{\rm and}~|x|~{\rm not}~{\rm
very}~{\rm small} \end{split} \ee
so that the related potentials $V_{R\pm}(x)\equiv \sigma_R^2 \pm
\sigma_R' (x)$ and $V_{L\pm}(x)\equiv \sigma_L^2 \pm \sigma_L'
(x)$ may have the general shapes shown in Figs.1 and 2,
respectively. Then the superpotential appropriate to our case is
obtained by combining these two types of functions as
\be\label{sp} W(x)=\sigma_R (x-l_1)+\sigma_L (x-l_2) +v \ee
with $L\equiv|l_1 -l_2|$ taken to be reasonably large (so that
$W(x)$ may have a {\it flat} basin between the points $x=l_2$ and
$x=l_1$).\footnote{At the later stage of our discussion we will
use the fact if $l^*$ denotes a certain point in the flat basin,
the approximation $\sigma_R (x-l_1)+v \simeq 0$ for $x<l^*$ or
$\sigma_L(x-l_2)+v \simeq 0$ for $x>l^*$ is valid.} See Fig.3 for
the schematic plots of $W(x)$ and the related potentials $V_{\pm}
(x) \equiv W^2 (x) \pm W'(x)$. Both $W(\infty)$ and $W(-\infty)$
being positive, this corresponds to the case of broken SUSY
\cite{Witt,Junk}; but, for the posited superpotential (with $L$
large), the ground state energy $\bar E$ is expected to be rather
small. [Our superpotential will be an even function of $x$ if
$\sigma_L (x)$ happens to be the mirror image of $\sigma_R (x)$,
i.e., $\sigma_L (x)=\sigma_R (-x)$, and $(l_1 ,l_2)$ are equal to
$(\frac{L}{2},-\frac{L}{2})$].

For $W(x)$ specified as above, the corresponding Hamiltonians
$H_{\pm}$ involve the potentials which can approximately be
described by the sum of two well-separated potentials (aside from
a constant term), i.e.,
\be \label{v} V_{\pm}(x) \sim V_{R\pm}(x-l_1) + V_{L\pm}(x-l_2) -
v^2 ~.\ee
These correspond to asymmetric double wells even when $W(x)$ is an
even function, and hence the well-known approximation schemes used
for symmetric double wells (e.g., instanton methods, tight-binding
approximations) would not be much useful. [Note that,
for the ground state of our Hamiltonian with the potential $V_+
\sim V_{R+} + V_{L+} -v^2$ (or $V_- \sim V_{R-} + V_{L-} -v^2$),
the tight-binding approximation is plainly not available --- while
the local Hamiltonian involving $V_{L+}$ (or $V_{R-}$) allows a
zero-energy bound state, no zero-energy state exists for the local
Hamiltonian involving $V_{R+}$ (or $V_{L-}$)]. But, for a
supersymmetric system, one can always consider a pair of coupled
first-order differential equations instead of the second-order
Schr\"odinger equations. Our perturbative approach for the ground
state is based on the analysis of these first-order equations, and
as a result we obtain a
remarkably simple formula
for the lowest eigenvalue $\bar E$. It is simply the {\it square}
of the product of the two zero-energy eigenfunctions (allowed with
the potentials $V_{L+} (x-l_2)$ and $V_{R-} (x-l_1)$
separately)
, evaluated at an
arbitrary chosen point $l^*$ in the flat middle region of the
superpotential $W(x)$. See the expression (\ref{18}) below.

Consider a matrix Hamiltonian
\be \label{mat} {\cal
H}=\begin{pmatrix}0&A^\dag\\A&0\end{pmatrix}\ee
with
\be A= \partial_x + W(x)~~~,~~~A^\dag = -\partial_x +W(x)~.\ee
The corresponding eigenvalue equation, ${\cal H} \Psi (x)=\omega
\Psi (x)$ with $\Psi (x)=\begin{pmatrix} \Psi_1 (x) \\ \Psi_2 (x)
\end{pmatrix}$, reduces to a pair of first-order differential equations
\begin{subeqnarray} \slabel{d1} A~\Psi_1 (x)= \omega \Psi_2 (x)~, \\
\slabel{d2} A^\dag ~\Psi_2 (x)= \omega \Psi_1 (x)~.
\end{subeqnarray}
Then, based on the relations
\be A^\dag A = -\partial^2_x + W^2 (x) -W'(x)\equiv H_- ~~~,~~~
AA^\dag = -\partial^2_x + W^2 (x) +W'(x)\equiv H_+ ~,\ee
one finds that the functions $\Psi_1(x)$ and $\Psi_2(x)$ are
eigenfunctions of the Schr\"odinger Hamiltonians $H_-$ and $H_+$
(with the same energy $E=\omega^2$), respectively. Further, it is
not difficult to show that if $\Psi (x) = \begin{pmatrix} \Psi_1
(x)
\\ \Psi_2 (x)
\end{pmatrix}$ is an eigenvector of ${\cal H}$ with eigenvalue
$\omega$, $\tilde\Psi (x) = \begin{pmatrix} \Psi_1 (x) \\ -\Psi_2
(x) \end{pmatrix}$ corresponds to an eigenvector of ${\cal H}$
with eigenvalue $-\omega$. For our later application, we also
remark that the above matrix Hamiltonian can be written as
\be \label{drc}{\cal H} = -i\gamma^0 \gamma^1 \partial_x +
\gamma^0 W(x)
\ee
with $\gamma^0 =\sigma^1 =\begin{pmatrix} 0 & 1 \\ 1 & 0
\end{pmatrix}$ and $\gamma^1 = i\sigma^3 = \begin{pmatrix} i & 0
\\ 0 & -i \end{pmatrix}$. That is, it describes the Dirac
Hamiltonian (defined on a line) in the presence of an external
scalar field $W(x)$ \cite{Coop}.

Let us now specialize to the case with the superpotential given by
the form (\ref{sp}) with $L$ large. If $\varphi_1(x)$ is the
ground-state eigenfunction of the Hamiltonian $H_-$ with a small
energy $\bar E\equiv \bar\omega^2$ and $\varphi_2(x)$ that of the
isospectral partner $H_+$ with the same energy, they will have to
satisfy the integral equations (derived from  Eqs. (\ref{d1}) and
(\ref{d2}))
\begin{subeqnarray} \slabel{19a}\varphi_1 (x) = e^{-\int^{x}_{l_1} W(y)dy}
\left\{D_1 + \bar\omega \int^{x}_{\alpha_1} dy
\left[e^{\int^y_{l_1} W(z)dz} \varphi_2(y)\right] \right\} ~, \\
\slabel{19b} \varphi_2(x)=e^{\int^{x}_{l_2} W(y)dy} \left\{D_2 -
\bar\omega \int^{x}_{\alpha_2} dy \left[e^{-\int^y_{l_2} W(z)dz}
\varphi_1(y)\right]\right\} ~,
\end{subeqnarray}
for suitable constants $\alpha_1$, $\alpha_2$, $D_1$ and $D_2$.
Clearly, for small enough $\bar\omega$, we may iterate these
equations to learn about the corresponding eigenfunctions. Then,
let $\varphi^0_1 (x)$ and $\varphi_2^0(x)$ denote the limiting
expressions of $\varphi_1(x)$ and $\varphi_2(x)$ for very large
$L$ (and therefore very small $\bar\omega$). Here, considering the
general shapes of the potentials $V_\pm$ (see also Eq.(\ref{v}))
shown in Fig.3, $\varphi^0_1(x)$ ($\varphi^0_2 (x)$) can be taken
as the normalized zero-energy eigenfunction of $-\partial_x^2
+V_{R-}(x-l_1)$ ($-\partial_x^2 +V_{L+}(x-l_2)$). This in turn
implies that $\varphi^0_1(x)$ and $\varphi^0_2 (x)$ satisfy the
first-order differential equations
\be\label{15} [\partial_x +
\sigma_R(x-l_1)]\varphi^0_1(x)=0~~~~~,~~~~~ [-\partial_x +
\sigma_L(x-l_2)]\varphi^0_2(x)=0~.\ee %
Hence we have the explicit expressions
\be \label{16} \varphi^0_1(x)= C_1 e^{-\int^x_{l_1} \sigma_R (y-
l_1)dy}~~~~~,~~~~~\varphi^0_2(x)= C_2 e^{\int^x_{l_2} \sigma_L (y-
l_2)dy}~\ee
with appropriate normalization constants $C_1$ and $C_2$, which
are taken to be positive. [In view of Eqs. (\ref{d1}) and
(\ref{d2}), both $\varphi^0_1(x)$ and $\varphi^0_2(x)$ are chosen
to be real]. As these informations are used with our integral
equations (\ref{19a}) and (\ref{19b}), we are led to the following
approximate expressions for $\varphi_1(x)$ and $\varphi_2(x)$
(valid for large $L$):
\begin{subeqnarray} \slabel{20a}\varphi_1 (x) \simeq e^{-\int^{x}_{l_1} W(y)dy}
\left\{C_1 + \bar\omega \int^{x}_{l^*} dy
\left[e^{\int^y_{l_1} W(z)dz} \varphi_2^0(y)\right] \right\} ~, \\
\slabel{20b} \varphi_2(x) \simeq e^{\int^{x}_{l_2} W(y)dy}
\left\{C_2 - \bar\omega \int^{x}_{l^*} dy \left[e^{-\int^y_{l_2}
W(z)dz} \varphi_1^0(y)\right]\right\}~~
\end{subeqnarray}
with the value $l^*$ chosen conveniently at some point in the flat
middle region of the superpotential so that both $|l_1-l^*|$ and
$|l^*-l_2|$ may become ${\cal O}(L)$.

We then note that the expression (\ref{20a}) as $x\rightarrow
-\infty$ (and similarly that in Eq.(\ref{20b}) as
$x\rightarrow\infty$) would blow up unless the value of
$\bar\omega$ were chosen such that
\be \label{eig}C_1 + \bar\omega \int^{-\infty}_{l^*}
dy\left[e^{\int_{l_1}^y W(z)dz} \varphi_2^0 (y) \right] = 0 ~.\ee
Here, with the explicit expressions for $W(x)$ and
$\varphi^0_2(x)$ given in Eqs. (\ref{sp}) and (\ref{16}), we
observe that
\be \begin{split} e^{\int^y_{l_1} W(z)dz} & \simeq
e^{\int^{l^*}_{l_1} \sigma_R (z-l_1)dz} e^{\int^{l_2}_{l^*}
\sigma_L (z-l_2)dz} e^{\int^{y}_{l_2} \sigma_L (z-l_2)dz} \\
&= C_1 [\varphi^0_1(l^*)\varphi^0_2(l^*)]^{-1}
\varphi^0_2(y)~~,~~~({\rm for}~y\in [-\infty,l^*])~,\end{split}
\ee
\be \int^{-\infty}_{l^*}dy[\varphi^0_2(y)]^2 \simeq
-\int^{\infty}_{-\infty}dy[\varphi^0_2(y)]^2
=-1~,~~~~~~~~~~~~~~~~~~~~~\ee
and hence the `eigenvalue condition' (\ref{eig}) reduces to the
form
\be C_1 -\bar\omega C_1 [\varphi^0_1(l^*)\varphi^0_2(l^*)]^{-1}
=0~.\ee
We thus find
\be \label{w} \bar\omega = \varphi^0_1(l^*)\varphi^0_2(l^*) = C_1
C_2 e^{-\int^{l^*}_{l_1} \sigma_R (y-l_1)dy + \int^{l^*}_{l_2}
\sigma_L (y-l_2)dy} \ee
and so, for the desired ground-state energy, the formula
\be\label{18}\bar E=[\varphi^0_1(l^*)\varphi^0_2(l^*)]^2 ~.\ee
[Notice that, for an arbitrary choice of $l^*$ in the flat middle
region of the superpotential, the same value for $\bar E$
results]. The same expression for $\bar E$ is obtained by
analogous analyses with Eq.(\ref{20b}). With the eigenvalue
$\bar\omega$ determined in this manner, the corresponding wave
functions $\varphi_1 (x)$ and $\varphi_2 (x)$ (up to
normalization) are now expressed as
\be\label{25} \varphi_1(x)\simeq \begin{cases}
\varphi_1^0(x)\left\{ 1+ \bar\omega \int^{x}_{l^*}dy
[\varphi^0_1(y)]^{-1}\varphi^0_2(y)\right\}~,& x>l^* \\
\bar\omega [\varphi^0_2(x)]^{-1}
\int^{x}_{-\infty}dy[\varphi^0_2(y)]^2~,& x<l^* ~~\end{cases} \ee
\be \label{26}\varphi_2(x) \simeq \begin{cases}
\bar\omega[\varphi^0_1(x)]^{-1}\int^{\infty}_{x}dy[\varphi^0_1(x)]^2~,&x>l^*\\
\varphi^0_2(x)\left\{1+\bar\omega\int^{l^*}_x
dy[\varphi^0_2(y)]^{-1} \varphi^0_1(y)
\right\}~,&x<l^*~.\end{cases}
\ee

We expect that a judicious use of the tight-binding approximation
with the {\it matrix Hamiltonian} (\ref{mat}), taking
$\begin{pmatrix}\varphi^0_1(x) \\ 0 \end{pmatrix}$ and
$\begin{pmatrix} 0 \\ \varphi^0_2(x) \end{pmatrix}$ as the
degenerate (i.e., zero energy) eigenstates of the corresponding
local Hamiltonians, lead to the same conclusion as
above.\footnote{But, since our matrix Hamiltonian (which is the
Dirac Hamiltonian) is unbounded from below, some care must be
exercised.} This is supported by the observation that, for the
eigenvalue $\bar\omega$, the same result (i.e., Eq.(\ref{w}))
follows from the calculation based on the formula
\be \label{27}\pm\bar\omega = \int^{\infty}_{-\infty}dx~
\varphi^0_\pm (x) {\cal H} \varphi^0_\pm (x)~,\ee
taking $\varphi^0_\pm
(x)=\frac{1}{\sqrt2}\begin{pmatrix}\varphi^0_1(x) \\
\pm \varphi_2^0(x)\end{pmatrix}$, the usual zeroth-order states in
the tight-binding approximation. Indeed, one can verify that the
given integral
\be\label{tb}\int^\infty_{-\infty}dx~\varphi^0_+(x){\cal H}
\varphi^0_+(x)=\int^{\infty}_{-\infty}dx ~[\sigma_R(x-l_1)
+v]\varphi_1^0(x)\varphi^0_2(x)\ee
is well approximated by the product
$\varphi^0_1(l^*)\varphi^0_2(l^*)$.

As an explicit example, consider the superpotential of the form
\be W(x)=v ~{\rm sgn}(x-\frac{L}{2}) - v ~{\rm sgn}(x+\frac{L}{2})
+v~,\ee
i.e., in our notation, $(l_l,l_2)=(\frac{L}{2},-\frac{L}{2})$,
$\sigma_R(x-l_1)=v ~{\rm sgn}(x-\frac{L}{2})$, and
$\sigma_L(x-l_2)=-v ~{\rm sgn}(x+\frac{L}{2})$. Given this, the
potentials of the Schr\"odinger Hamiltonians $H_\pm$ will be
%
\be V_\pm(x)=v^2 \pm 2v~\delta (x-\frac{L}{2}) \mp 2v ~\delta
(x+\frac{L}{2})~.\ee
For these systems, one can of course find the exact ground state
energy $\bar E$ by solving the appropriate Schr\"odinger
equations. This exercise shows that $\bar E$ is the root of the
equation $\bar E= v^2 e^{-2L{\sqrt{v^2-\bar E}}}$, and hence, for
large $L$, we have
\be \label{31}\bar E \simeq v^2 e^{-2vL}~.\ee
Let us see whether our formula (\ref{18}) yields the same. The
normalized solutions of Eq.(\ref{15}) are trivially found here:
\be \varphi^0_1(x)={\sqrt v}~e^{-v|x-\frac{L}{2}|}~~~~~,~~~~~
\varphi^0_2(x)={\sqrt v}~e^{-v|x+\frac{L}{2}|}~.\ee
Then, from Eq.(\ref{18}), we have that $\bar E$ (for large $L$)
should equal $[\varphi^0_1(0)\varphi^0_2(0)]^2 = v^2 e^{-2vL}$.
Hence a complete agreement.

More physically relevant example is provided by the Dirac
Hamiltonian (\ref{drc}) with the scalar field taken to represent
the kink-antikink pair,
\be \label{33}W(x)= v\tanh\left[\frac{\mu}{2}
(x-\frac{L}{2})\right] - v\tanh\left[\frac{\mu}{2}
(x+\frac{L}{2})\right] +v~.\ee
Here the scalar field $\sigma_R(x-\frac{L}{2}) = v\tanh
\left[\frac{\mu}{2} (x-\frac{L}{2})\right]$ represents a kink
located at $x=\frac{L}{2}$, and $\sigma_L(x+\frac{L}{2}) = -v\tanh
\left[\frac{\mu}{2} (x+\frac{L}{2})\right]$ an antikink at
$x=-\frac{L}{2}$ \cite{Soli}. The widely-separated kink-antikink
configuration, described by the form (\ref{33}), has received
attention in Refs. \cite{Jack,Schon,Alt}. Especially interesting
is the almost zero-energy mode of the Dirac Hamiltonian, in
connection with the role of the so-called Jackiw-Rebbi mode
\cite{Jack} (which refers to the zero-energy fermion mode
\cite{Choon} in the kink or antikink background) when a kink and
an antikink are simultaneously present. In the kink or antikink
background the Jackiw-Rebbi mode is represented by
$\begin{pmatrix}\varphi^0_1(x) \\ 0
\end{pmatrix}$ or $\begin{pmatrix} 0 \\ \varphi^0_2(x)
\end{pmatrix}$, if $\varphi^0_1(x)$ and $\varphi^0_2(x)$ denote
the normalized solutions of Eq.(\ref{15}):
\be
\varphi^0_1(x)=C\left[\cosh\frac{\mu}{2}(x-\frac{L}{2})\right]^{-\frac{2v}{\mu}}
~~~~~,~~~~~
\varphi^0_2(x)=C\left[\cosh\frac{\mu}{2}(x+\frac{L}{2})\right]^{-\frac{2v}{\mu}}
~,\ee
where $C=\left\{
\frac{\mu}{2}\frac{\Gamma(\frac{2v}{\mu}+\frac{1}{2})}
{\Gamma(\frac{2v}{\mu})\Gamma(\frac{1}{2})}\right\}^{\frac{1}{2}}$.
Then, in the above kink-antikink scalar background, one can
immediately find the energy of the almost-zero-energy fermion
eigenmode by using our formula (\ref{w}) --- it is equal to $\pm
\bar\omega$, with\footnote{Based on some heuristic arguments the
author of Ref.\cite{Alt} also identified $\bar\omega$ with the
expectation value in Eq.(\ref{27}). But his final expression for
$\bar\omega$ is apparently not consistent with our result in
Eq.(\ref{35})---this is due to the calculational error made in
Ref.\cite{Alt} (e.g., wrong fermion zero-energy wave functions
used).}
\be \label{35}\begin{split} \bar\omega &=
\varphi^0_1(0)\varphi^0_2(0)=
C^2[\cosh\frac{\mu L}{4}]^{-\frac{4v}{\mu}} \\
&\simeq \frac{\mu}{2}e^{\frac{4v}{\mu}}
\frac{\Gamma(\frac{2v}{\mu}+\frac{1}{2})}
{\Gamma(\frac{2v}{\mu})\Gamma(\frac{1}{2})} e^{-vL}~.
\end{split}\ee
%
The exponential dependence
of $\bar\omega$ on the distance $L$ was previously noted in
Ref.\cite{Jack}. This $L$-dependent fermion energy shift will
contribute to the effective potential between the kink and the
antikink. For instance, in the vacuum sector where all negative
energy fermion modes are to be occupied, the contribution from
this mode, i.e., that with energy $-\bar\omega$ will become more
negative as $L$ decreases, thus producing an attractive
interaction between the kink and the antikink (if only this mode
is taken into account).

In this work we investigated on some special properties pertaining
to the ground state of a quantum mechanical Hamiltonian with
broken supersymmetry, when the corresponding eigenvalue is small.
A direct perturbative analysis based on the first-order coupled
equaitons has been used to obtain a very simple (approximate)
expression for the ground state energy. Our formula (\ref{18})
should be useful in finding the ground state energy of a
Schr\"odinger Hamiltonian the potential in which can be
approximated by the form (\ref{v}).

\section*{Acknowledgment}
We would like to thank Seok Kim for very useful discussions. This
work was supported in part by the BK21 project of the Ministry of
Education, Korea, and also by Korea Research Foundation Grant
2001-015-DP0085(C.L.).

\newpage
\begin{figure} 
\centering\psfrag{sr}{$\sigma_R$}\psfrag{v2}{$v^2$}\psfrag{-v}{$-v$}
\psfrag{x}{$x$} \psfrag{V+}{$V_{R+}(x)$} \psfrag{V-}{$V_{R-}(x)$}
\psfrag{VR}{$V_R$}
\includegraphics[clip=,width=.8\textwidth]{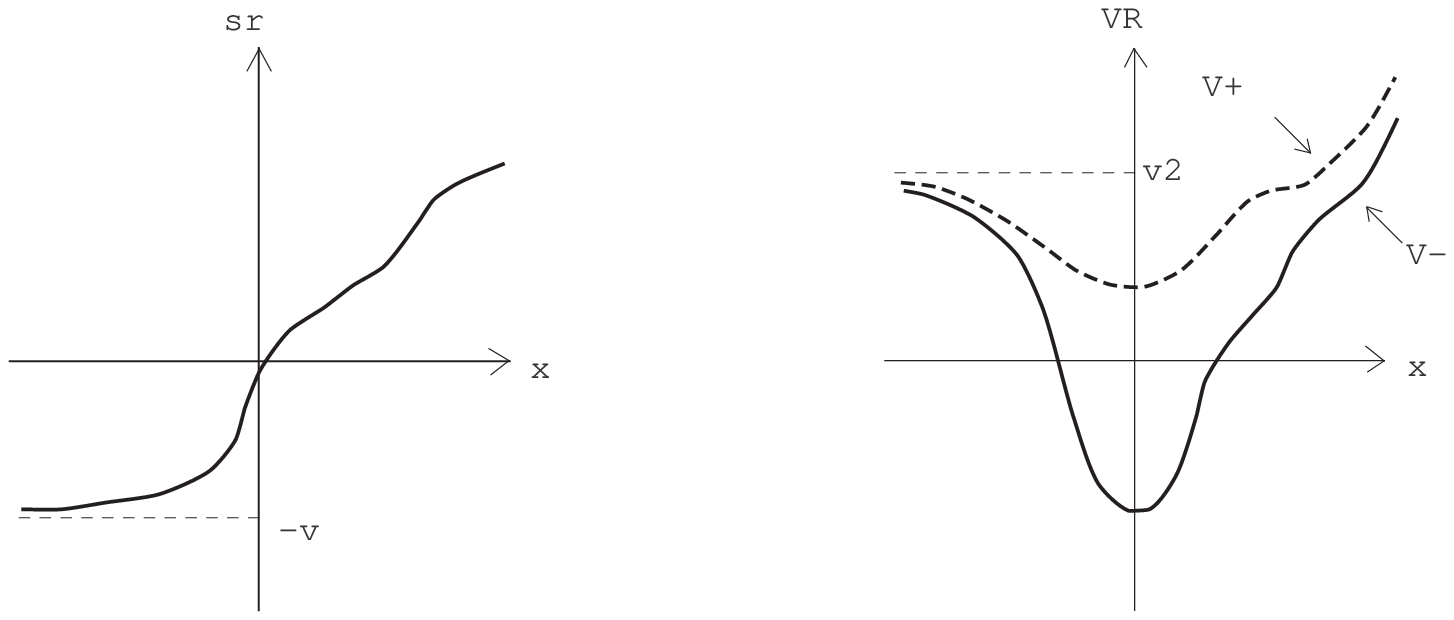}
\caption{\footnotesize{Schematic plots of $\sigma_R (x)$ and
$V_{R\pm}(x) \equiv \sigma_R^2 (x) \pm \sigma_R' (x)$}}
\end{figure}
\begin{figure} 
\centering\psfrag{sl}{$\sigma_L$}\psfrag{v2}{$v^2$}\psfrag{-v}{$-v$}
\psfrag{x}{$x$} \psfrag{V+}{$V_{L+}(x)$} \psfrag{V-}{$V_{L-}(x)$}
\psfrag{VL}{$V_L$}
\includegraphics[clip=,width=.8\textwidth]{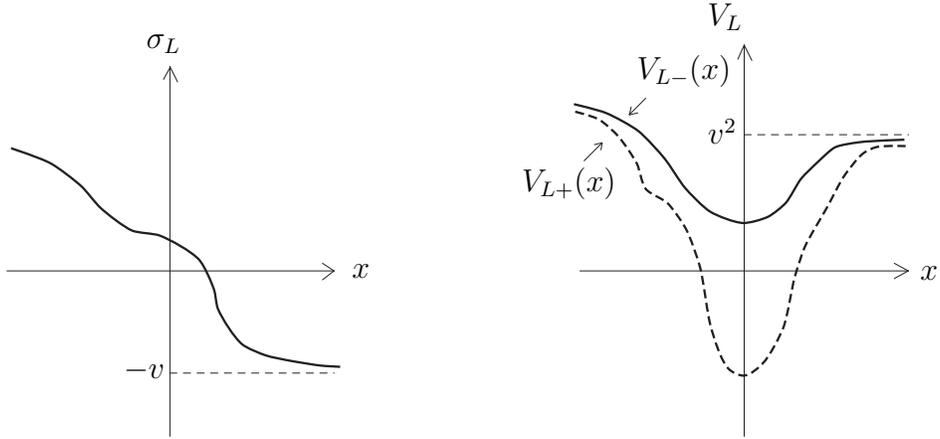}
\caption{\footnotesize{Schematic plots of $\sigma_L (x)$ and
$V_{L\pm}(x) \equiv \sigma_L^2 (x) \pm \sigma_L' (x)$}}
\end{figure}

\newpage
\begin{figure} 
\centering\psfrag{W}{$W$}\psfrag{v2}{$v^2$}
\psfrag{l1}{$l_1$}\psfrag{l2}{$l_2$}\psfrag{ls}{$l^*$}
\psfrag{x}{$x$} \psfrag{V+}{$V_{+}$} \psfrag{V-}{$V_{-}$}
\includegraphics[clip=,width=1.0\textwidth]{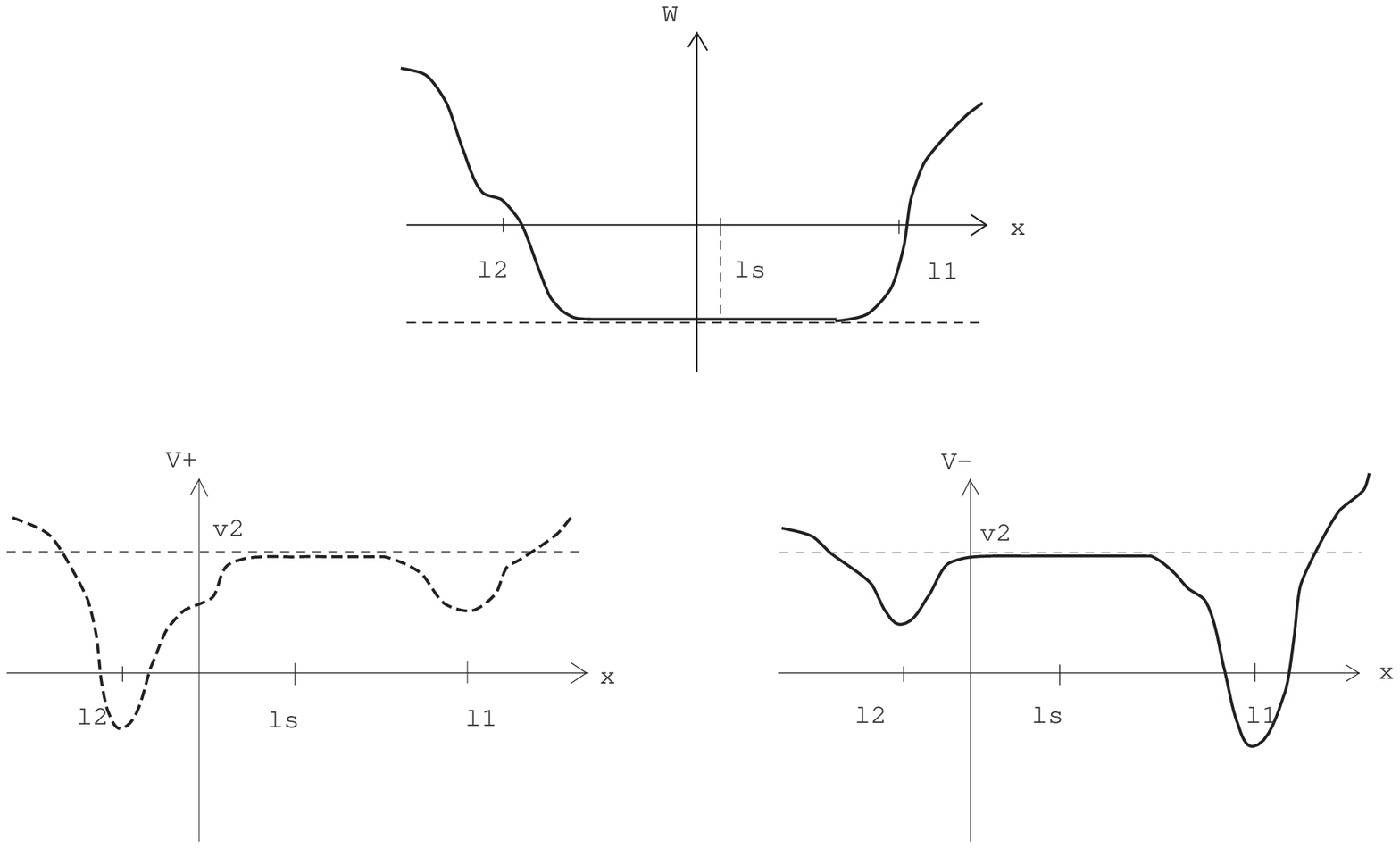}
\caption{\footnotesize{Schematic plots of $W(x)$ and $V_\pm (x)
\equiv W^2 (x) \pm W'(x)$}}
\end{figure}
%


\begin{thebibliography}{99}
\bibitem{Witt} E.~Witten, Nucl.\ Phys.\ {\bf B188}, 513 (1981);
F.~Cooper and B.~Friedman, Ann.\ Phys.\ (NY) {\bf 146}, 262
(1983).
\bibitem{Junk} G.~Junker, Supersymmetric Methods in Quantum
and Statistical Physics,\ Springer, Berlin, 1996.
\bibitem{Gend} L.~Gendenshtein, JETP Lett. {\bf 38}, 356 (1983);
L.~Gendenshtein and I.~V.~Krive, Sov.\ Phys.\ Usp.\ {\bf 28}, 645
(1985).
\bibitem{Dutt} R.~Dutt, A.~Khare, and U.~Sukhatme, Phys.\ Lett.\ {\bf
B181}, 295 (1986); R.~Dutt, A.~Khare, and U.~Sukhatme, Am.\ J.\
Phys.\ {\bf 56}, 163 (1988).; C.~V.~Sukumar, J.\ Phys.\ {\bf A18},
2917 (1985).
\bibitem{CKS} F.~Cooper, A.~Khare, and U.~Sukhatme, Phys.\ Rep.\ {\bf
251}, 268 (1995).
\bibitem{Dutt2} R.~Dutt, A.~Gangopadhyaya, A.~Khare,
A.~Pagnamenta, and U.~Sukhatme, Phys.\ Lett.\ {\bf A174}, 364
(1993).
\bibitem{Gang} A.~Gangopadhyaya, J.~V.~Mallow, and U.~P.~Sukhatme,
Phys.\ Lett.\ {\bf A283}, 279 (2001).
\bibitem{Coop} F.~Cooper, A.~Khare, R.~Musto, and A.~Wipf, Ann.\ Phys.\ {\bf
187}, 1 (1988).
\bibitem{Soli} R.~Rajaraman, Solitons and Instantons,
North-Holland, 1982.
\bibitem{Jack} R.~Jackiw and C.~Rebbi, Phys.\ Rev.\ {\bf D13}, 3398 (1976).
\bibitem{Schon} J.~F.~Schonfeld, Nucl.\ Phys.\ {\bf B161}, 125 (1979);
A.~S.~Goldhaber, A.~Litvintsev, and P.~van Nieuwenhuizen, Phys.\
Rev.\ {\bf D64}, 045013 (2001).
\bibitem{Alt} B.~Altschul, hep-th/0111042.
\bibitem{Choon} C.~Lee, unpublished;
R. Dashen, B.~Hasslacher, and A.~Neveu, Phys.\ Rev.\ {\bf D10},
4130 (1974); W.~Bardeen, M.~Chanowitz, S.~Drell, M.~Weinsrein, and
T.~-M.~Yan, Phys.\ Rev.\ {\bf D11}, 1094 (1975).
\end{thebibliography}
\end{document}